\documentclass[aps,prl,twocolumn,superscriptaddress]{revtex4-2}
\usepackage{amsmath}
\usepackage{mathtools}
\usepackage{amssymb}
\usepackage{mathrsfs}
\usepackage{dsfont}
\usepackage[pdftex,colorlinks=true,urlcolor=blue,linkcolor=blue,citecolor=blue,breaklinks=true,bookmarks=false]{hyperref}
\usepackage{graphicx}
\usepackage{float}
\usepackage{xcolor}

\usepackage{tikz}
\begin{document}

\title{\textit{Anti-thermalization:} Heating by cooling and vice versa}

\author{Jaswanth Uppalapati}
\affiliation{Indian Institute of Science, Bangalore 560012, India }
\author{Paul A McClarty}
\affiliation{Laboratoire Léon Brillouin, CEA, CNRS, Université Paris-Saclay, CEA Saclay, 91191 Gif-sur-Yvette, France.}
\affiliation{Max Planck Institute for the Physics of Complex Systems, N\"{o}thnitzer Str. 38, 01187 Dresden, Germany}
\author{Masudul Haque}
\affiliation{Institut f\"ur Theoretische Physik, Technische Universit\"at
Dresden, 01062 Dresden, Germany}
\affiliation{Max Planck Institute for the Physics of Complex Systems, N\"{o}thnitzer Str. 38, 01187 Dresden, Germany}
\author{Shovan Dutta}
\email[E-mail: ]{shovan.dutta@rri.res.in}
\affiliation{Raman Research Institute, Bangalore 560080, India}


\begin{abstract}
Common intuition tells us that if one part of a connected system is cooled continuously, the other parts should also cool down. This intuition can be given a microscopic foundation for the case of a generic quantum system coupled to a ``lead'' that is maintained at a given temperature. We show that by suppressing resonant energy exchange between the two parts, one can reverse the fate of the system, namely, it can heat up toward its most excited state as the lead is cooled to its ground state, and vice versa. This anti-thermal dynamics arises in a broad class of systems with a conserved $U(1)$ charge, and can be tested with two qubits in existing setups. We show that the mechanism allows one to prepare mid-spectrum nonclassical states, stable temperature gradients in closed systems, and highly athermal states where subspaces heat in the presence of overall cooling. Our findings highlight the critical role played by the nature of the coupling and reveal a rich interplay between symmetry and resonance effects in the dynamics of thermalization.
\end{abstract}

\maketitle

\emph{Introduction.}|Central to statistical mechanics is the idea that a generic system, part of which is coupled to a heat bath, equilibrates to the bath temperature \cite{Pathria_statmech, Schieve_quantum_statmech}. This forms the foundation for sympathetic cooling \cite{Larson1986, Myatt1997, Kielpinski2000, Modugno2001, Bohman2021, Mao_etal_PRL2021_SympatheticCooling_TrappedIonCrystal}, a key experimental protocol underlying many of the successes in preparing synthetic quantum matter. Building on this idea and motivated by growing experimental capabilities to engineer local dissipation \cite{Poyatos1996, Barreiro2011, Morigi2015, Ma2019, Mendona2020, Li2023}, several recent studies have modeled the evolution of an extended quantum system coupled locally to an auxiliary system or ``lead'' that is actively maintained at a given temperature \cite{Cormick2013, Reichental2018, Palmero2019, Raghunandan2020, Tupkary2023, Zanoci2023, Langbehn2024}. It was found that even a single-site, zero-temperature lead can efficiently cool down many-body quantum systems \cite{Raghunandan2020, Langbehn2024}. Thermalization to the lead temperature improves as the lead size is increased \cite{Zanoci2023} and becomes exact for an infinite free-fermion lead with Markovian dissipation and large bandwidth in the limit of slow cooling and weak system-lead coupling \cite{Reichental2018}.

Here, we show that the above framework relies critically on the generic nature of the coupling---a simple but structured coupling can produce dramatic scenarios where the system heats up toward its most excited state as the lead is cooled to its ground state and vice versa. As we discuss below, the mechanism behind the usual sympathetic cooling is resonant energy exchange between the system and the lead \cite{Raghunandan2020}. By suppressing these energy-conserving processes, one can reverse the temperature of the system relative to the lead, leading to a stable temperature gradient even after the dissipation is turned off. 

We show how this ``anti-thermalization'' can be realized in current experiments by harnessing a conserved $U(1)$ charge and local, on-site fields for a broad class of systems, both integrable and ergodic. By tuning the local fields one can induce a competition between heating and cooling in the system and stabilize mid-spectrum, nonclassical states (e.g., an $N$-photon Fock state) or highly athermal states where the system cools overall but heats within each symmetry sector (or the other way around). These predictions do not require Markovian dissipation and survive up to finite symmetry-breaking terms.

\emph{Basic mechanism.}|Before presenting the anomalous behavior, let us first understand how sympathetic cooling works by considering a single-qubit lead with Markovian dissipation. Suppose a finite system with energy levels $\{|E_i\rangle\}$ is coupled to the lead with energy levels $\{ |\!\uparrow\rangle, |\!\downarrow\rangle \}$ and gap $\Delta_{\rm L} = \varepsilon_{\downarrow} - \varepsilon_{\uparrow} > 0$. The coupling is $\epsilon \smash{\hat{H}_{\rm SL}}$ and the cooling of the lead is described by a Lindblad operator \cite{Breuer_Lindblad} $\smash{\hat{L}} = \sqrt{\gamma} \hat{\sigma}^+$, where $\gamma$ is the cooling rate. We focus on small $\epsilon$ and $\gamma$, when the broadening of the energy levels is minimal and the thermalization is most accurate. In the absence of degeneracies, the eigenstates of the full Hamiltonian, to first order in $\epsilon$, are of the form
\begin{equation}
    |\uparrow, E_i \rangle_{\epsilon} = 
    |\uparrow, E_i \rangle 
    + \epsilon \sum_{j \neq i} c_{i,j} |\uparrow, E_j \rangle
    + \epsilon \sum_j d_{i,j} |\downarrow, E_j \rangle \;,
\end{equation}
where $|\sigma, E_i \rangle \coloneqq |\sigma\rangle \otimes |E_i\rangle$ for $\sigma = \uparrow, \downarrow$ and the coefficients $c_{i,j}$ and $d_{i,j}$ are given by perturbation theory. When $\gamma$ is weak, the density matrix is approximately diagonal in the energy basis and the populations evolve under Pauli rate equations due to transitions caused by $\hat{L}$ \cite{Vorberg_Ketzmerick_Eckardt_PRL2013_BEC_RateEqns, Vorberg_Schomerus_Ketzmerick_Eckardt_PRE2015, Wu2019, Liul_Shevchenko_LTP2023RateEquationQubits, Schnell_Widera_Eckardt_PRA2023_scarlike}. For $\gamma \gg \epsilon$ the lead is always close to its ground state, which gives the transition rates
\begin{equation}
    \Gamma_{i \rightarrow j} \approx 
    |{}_{\epsilon}\langle \uparrow, E_j | \hat{L} | \uparrow, E_i \rangle_{\epsilon}|^2 \approx
    \gamma \epsilon^2 \; 
    \frac{|\langle \downarrow, E_j | \hat{H}_{\rm SL} | \uparrow, E_i \rangle |^2}{(E_i - E_j - \Delta_{\rm L})^2} \;.
    \label{eq:single-qubit_rates}
\end{equation}
For a generic coupling, this rate diverges for $E_i - E_j = \Delta_{\rm L}$, when the energy reduction of the system is exactly compensated by exciting the lead. For a lead with infinite bandwidth, these processes resonantly cool the system to its ground state. More generally, the denominator favors cooling by amplifying transitions with $E_j < E_i$.

\begin{figure}[tbp]
    \centering
    \includegraphics[width=1\columnwidth]{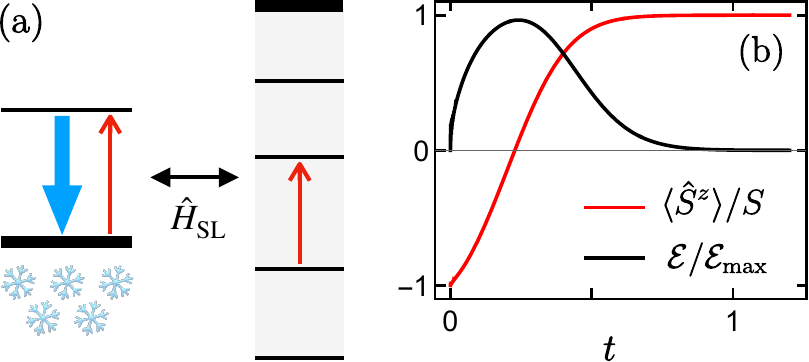}
    \caption{\label{fig:mechanism}(a) Mechanism of heating by cooling for a single-qubit lead (left) coupled to a multilevel system (right). The lead is continuously cooled to its ground state (blue arrow) while the coupling $\epsilon \smash{\hat{H}_{\rm SL}}$ excites both parts (red arrows). (b) Evolution of the normalized energy, $\smash{\langle\hat{S}_z\rangle/S}$, and entropy, $\mathcal{E}$, of a spin system $\smash{\hat{H}_{\rm S} = \hat{S}^z}$ with $S=10$ for spin-exchange coupling with $\epsilon = 0.2$ and lead Hamiltonian $\hat{H}_{\rm L} = -\hat{\sigma}^z$. The cooling is modeled by a Lindblad operator $\hat{L} = \sqrt{\gamma} \hat{\sigma}^+$ with $\gamma = 0.2$.
    }
\end{figure}

A key assumption in these arguments is that the coupling $\hat{H}_{\rm SL}$ has nonzero matrix elements for energy-conserving processes. Below we give simple examples for which this is not the case; instead, $\hat{H}_{\rm SL}$ either excites or de-excites both the system and the lead. If, in addition, the lead is continuously brought back to its ground state, the net effect is to drive the system up in energy until it cannot be excited any further by the coupling, as sketched in Fig.~\ref{fig:mechanism}(a). This holds irrespective of the spectral properties (e.g., level spacings) of the system. Note that the process does not violate the second law of thermodynamics as the temperature difference between the lead and the system is not spontaneously generated; rather, it requires work to actively cool the lead, which is \emph{necessary} for heating the system, as we explain further below. If the cooling is turned off at any stage, the coupling cannot redistribute energy between the two parts, so their temperature difference persists indefinitely.

\emph{Single-site lead.}|The simplest instance of heating by cooling occurs for a single-qubit lead coupled to an multi-level system which may represent a (collective) spin $S$. The physics is largely independent of $S$. Suppose the two spins are kept in opposite magnetic fields such that $\hat{H}_{\rm L} = -\Delta_{\rm L} \hat{\sigma}^z$ and $\hat{H}_{\rm S} = \Delta_{\rm S}\hat{S}^z$ with $\Delta_{\rm S}, \Delta_{\rm L} > 0$, and they have a spin-exchange coupling $\hat{H}_{\rm SL} = \hat{\sigma}^+ \hat{S}^- + \hat{\sigma}^- \hat{S}^+$. Such interactions arise naturally in a number of quantum simulator platforms \cite{Georgescu2014}, where the on-site potentials can be tuned individually. Since the energy varies oppositely with $\sigma^z$ and $S^z$ and the coupling conserves their sum, $\hat{H}_{\rm SL}$ either raises or lowers the energy of both spins. Therefore, if the lead always relaxes to its ground state, the system spin will reach its most excited state. This can be checked explicitly for Markovian cooling using a Lindblad operator, as shown in Fig.~\ref{fig:mechanism}(b). Starting from its ground state, the energy of the system rises monotonically and saturates to its maximum value, while the lead is cooled to its ground state. The entropy of the system reaches its peak at an intermediate time when it is maximally spread among all energy levels.

An apparent paradox here is how the cooling, which extracts energy from the lead, is able to pump energy into the system or increase the total energy. To understand this, consider the case where the system is also a qubit. Then, to first order in $\epsilon$, all eigenstates of the full Hamiltonian can be sketched as follows [with $\tilde{\epsilon} \coloneqq \epsilon/(\Delta_{\rm L} + \Delta_{\rm S})$].
\begin{equation}
\begin{tikzpicture}[line width=1pt, baseline=(current bounding box.center)]

\draw (-2.1, 0) -- (-0.7, 0); 
\node[left] at (-2.1, 0.1) {$-\frac{\Delta_{\rm L} + \Delta_{\rm S}}{2}$};
\node[right] at (-0.7, 0) {$|\!\uparrow \downarrow\rangle - \tilde{\epsilon} |\!\downarrow \uparrow\rangle$};

\draw[dashed] (-2.1, 0.2) -- (-0.7, 0.2); 
\node[above] at (-1.5, 0.2) {$|\!\uparrow\downarrow\rangle$};

\draw[dashed] (-2.1, 1.8) -- (-0.7, 1.8); 
\node[below] at (-1.5, 1.85) {$|\!\downarrow\uparrow\rangle$};

\draw (-2.1, 2) -- (-0.7, 2); 
\node[left] at (-2.1, 1.9) {$\frac{\Delta_{\rm L} + \Delta_{\rm S}}{2}$};
\node[right] at (-0.7, 2) {$|\!\downarrow \uparrow\rangle + \tilde{\epsilon} |\!\uparrow \downarrow\rangle$};

\draw (1.4, 0.6) -- (2.8, 0.6); 
\node[right] at (2.8, 0.6) {$\frac{\Delta_{\rm S} - \Delta_{\rm L}}{2}$};
\node[below] at (2.1, 0.6) {$|\!\uparrow\uparrow\rangle$};

\draw (1.4, 1.4) -- (2.8, 1.4); 
\node[right] at (2.8, 1.4) {$\frac{\Delta_{\rm L} - \Delta_{\rm S}}{2}$};
\node[above] at (2.1, 1.4) {$|\!\downarrow\downarrow\rangle$};

\draw[red,->,>=stealth] (-1, 0) -- (-1, 1.75);
\draw[blue,->,>=stealth] (-0.9, 1.75) -- (2, 0.65);

\end{tikzpicture}
\label{eq:energy_levels}
\end{equation}
One way to implement the cooling is to measure $\hat{\sigma}^z$ and, if the outcome is $\downarrow$, flip the spin using a $\pi$ pulse \cite{Wiseman1994}. Thus, starting from the overall ground state, $|\!\uparrow \downarrow\rangle - \tilde{\epsilon} |\!\downarrow \uparrow\rangle$, the measurement can project the spins to $|\!\downarrow\uparrow\rangle$ (red arrow) with probability $\tilde{\epsilon}^2$, before they are cooled to the dark state $|\!\uparrow\uparrow\rangle$ (blue arrow). The first, nonunitary step increases the energy by $\Delta_{\rm L} + \Delta_{\rm S}$, while the second lowers it by $\Delta_{\rm L}$, resulting in an overall increase by $\Delta_{\rm S}$.

The anti-thermal behavior is not limited to ground-state cooling. The lead qubit can be held at an arbitrary inverse temperature $\beta_{\rm L}$ through a combination of pump and loss with relative rates $\gamma_+ / \gamma_- = e^{-\beta_{\rm L} \Delta_{\rm L}}$. Then the rates in Eq.~\eqref{eq:single-qubit_rates} give both upward and downward transitions between neighboring energy levels of the spin-$S$ system with $\Gamma_{\rm up} / \Gamma_{\rm down} = e^{\beta_{\rm L} \Delta_{\rm L}}$. As we show in Appendix \ref{app:detailed_balance}, in such cases the steady state can be found exactly using detailed balance. For a uniform level spacing $\Delta_{\rm S}$, this gives a thermal distribution with inverse temperature $\beta_{\rm S} = -(\Delta_{\rm L} / \Delta_{\rm S}) \beta_{\rm L}$.

The same arguments can be generalized to an extended system (e.g., an XXZ chain) for which the total $z$ projection $\hat{S}^z$ is a good quantum number. The lead qubit can be attached to one site of the system with a spin-exchange coupling, as before. If the qubit is continuously cooled to its ground state $|\!\uparrow\rangle$, the coupling will raise the spin of the coupled site, increasing $S^z$ by $1$. As long as the system Hamiltonian can spread this local excitation, the same site can be excited again, eventually driving the system to the state $|\!\uparrow \uparrow \dots \uparrow\rangle$. Depending on $\hat{H}_{\rm S}$ this state can lie anywhere in the spectrum, including at the very top for a sufficiently large magnetic field along $-z$.

\begin{figure}[tbp]
    \centering
    \includegraphics[width=1\columnwidth]{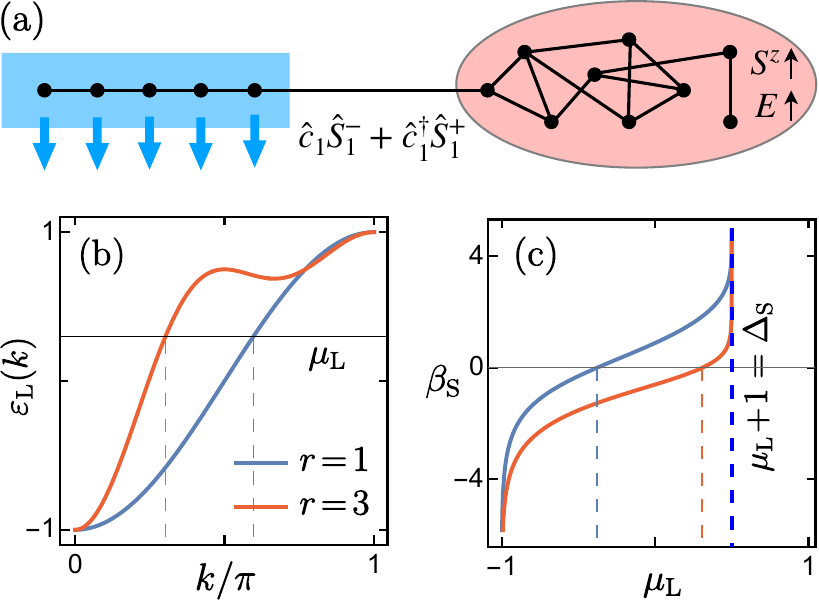}
    \caption{\label{fig:freefermion}(a) Free-fermionic lead (left) coupled to a spin system (right) whose energy increases with the total $z$ magnetization $\smash{\hat{S}^z}$. (b) Rescaled lead dispersion $\varepsilon_{\rm L} = -\sum_{m=1}^r \cos (m k) / m$, where $r$ is a hopping range. (c) Steady-state inverse temperature for the system $\smash{\hat{H}_{\rm S} = \Delta_{\rm S} \hat{S}^z}$ with $\Delta_{\rm S} = 1.5$ vs the lead chemical potential $\mu_{\rm L}$. Dashed lines in (b) and (c) show Fermi momenta and crossover from heating to cooling, respectively.
    }
\end{figure}

\emph{Free-fermion lead.}|Using a lead with more energy levels allows greater flexibility to shape the evolution of the system. To see this, consider a multisite free-fermion lead described by $\hat{H}_{\rm L} = -J_{\rm L} \sum_i (\hat{c}_i^{\dagger} \hat{c}_{i+1} + \text{h.c.}) - \mu_{\rm L} \sum_i \hat{c}_i^{\dagger} \hat{c}_i$, which may be realized as a spin-$1/2$ XY chain where $\mu_{\rm L}$ maps onto a uniform magnetic field \cite{Jordan1928}. Suppose one end of the chain is coupled to one site of an extended system, as sketched in Fig.~\ref{fig:freefermion}(a), for which $S^z$ is a good quantum number and the energy increases with $S^z$. After mapping the $\uparrow$ and $\downarrow$ spin states of the lead to empty and filled sites, respectively, the spin-exchange coupling takes the form $\hat{H}_{\rm SL} = \hat{c}_1 \hat{S}_1^- + \hat{c}_1^{\dagger} \hat{S}_1^+$. 

Let us focus on cooling the lead to its ground state, which is found by writing $\hat{H}_{\rm L} = \sum_k [\varepsilon_{\rm L}(k) - \mu_{\rm L}] \smash{\hat{\chi}_k^{\dagger} \hat{\chi}_k}$, with $\varepsilon_{\rm L} = -2 J_{\rm L} \cos k$ and $\hat{\chi}_k = \smash{[2/(l+1)]^{\frac{1}{2}}} \sum_i \sin(k i) \hat{c}_i$. Here $l$ is the number of sites and $k = q\pi/(l+1)$ with $q \in \{1,2,\dots,l\}$. In the ground state, all modes with $\varepsilon_{\rm L}(k) < \mu_{\rm L}$ are occupied and the rest are empty. In particular, for $\mu_{\rm L} < -2 J_{\rm L}$, all sites are empty and the coupling creates particle excitations through the term $\smash{\hat{c}_1^{\dagger} \hat{S}_1^+}$, which heats the system to maximal $S^z$. Note that this occurs even for an infinite lead, weak coupling, and slow cooling rate, which are ideal for sympathetic cooling \cite{Reichental2018}. As one sweeps $\mu_{\rm L}$ across the band, more and more modes are occupied and the system goes from being maximally heated to maximally cooled. This can be checked for Markovian cooling using one Lindblad operator for each mode, $\hat{L}_k = \sqrt{\gamma} \hat{\chi}_k^{\dagger}$ if $\varepsilon_{\rm L}(k) < \mu_{\rm L}$ and $\hat{L}_k = \sqrt{\gamma} \hat{\chi}_k$ otherwise, which give the transition rates (see Appendix \ref{app:detailed_balance})
\begin{equation}
    \frac{\Gamma_{i \rightarrow j}}{\gamma \epsilon^2} = \frac{2}{\pi}
    \sum_{p=\pm} \int_{p(\varepsilon_{\rm L}(k) - \mu_{\rm L}) > 0} \!
    \frac{{\rm d}k \; \sin^2 \! k \; |\langle E_j | \hat{S}_1^p | E_i \rangle|^2}{(E_i - E_j - |\varepsilon_{\rm L}(k) - \mu_{\rm L}|)^2} .
    \label{eq:free_fermion_rates}
\end{equation}
For a single-spin system the steady state can again be found using detailed balance. A uniform level spacing $\Delta_{\rm S}$ yields a thermal state that shows the expected crossover from heating to cooling, as in Fig.~\ref{fig:freefermion}(c) for $J_{\rm L} = 1/2$. 

\begin{figure}[tbp]
    \centering
    \includegraphics[width=1\columnwidth]{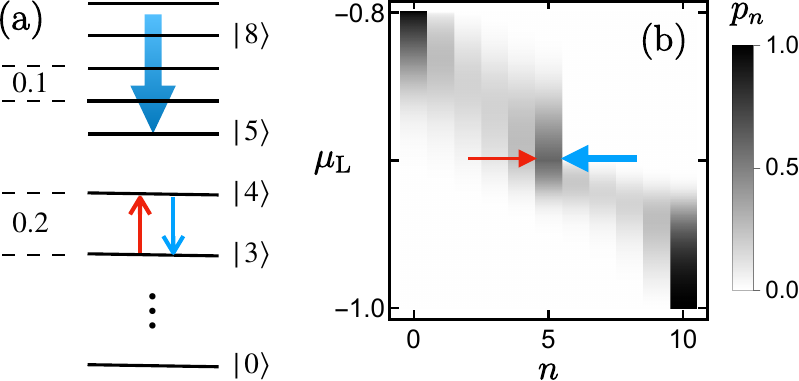}
    \caption{\label{fig:Fock}(a) Spectrum of a nonlinear cavity whose gap changes from $0.2$ to $0.1$ beyond the Fock state $n=5$. Coupling this system to the free-fermion lead with $0.1 < \mu_{\rm L} + 1 < 0.2$ gives resonant cooling above $n = 5$ and both heating and cooling below. (b) Steady-state distribution showing a peak at $n = 5$.
    }
\end{figure}

Note the system is completely cooled even before $\mu_{\rm L}$ crosses the band. This threshold occurs at $\mu_{\rm L} + 2 J_{\rm L} = \Delta_{\rm S}$ (for $\Delta_{\rm S} < 4 J_{\rm L}$) and signals the onset of sympathetic cooling. Beyond this point, the energy lowered by $\hat{S}_1^-$ can be fully compensated by a hole excitation in the lead. Figure~\ref{fig:freefermion} also shows that by including longer range hops in the lead, one can reduce the number of occupied modes for a given $\mu_{\rm L}$ and extend the region of negative system temperature. For a two-level system, the Lindblad dynamics can be solved exactly using third quantization \cite{Prosen2008}, which reproduces these results for large $l$, small $\gamma$, and small $\epsilon$ (Fig.~\ref{fig:third_quantization}). For large $\gamma$ one enters the quantum Zeno limit where the threshold disappears as resonances are broadened in a projected subspace, as formulated in Ref.~\cite{Popkov2018} and worked out in Appendix \ref{app:third_quantization}. The controlled competition of heating and cooling can be harnessed to stabilize more exotic states, as we discuss below.

\emph{Mid-spectrum nonclassical states.}|A long-standing challenge for nonlinear quantum optics is to generate nonclassical, number-squeezed photon states with sub-Poissonian statistics, which are of both fundamental and technological interest \cite{Davidovich1996, Hofheinz2008, Uria2020, Lingenfelter2021, Rivera2023}. Here we consider a simple model for a nonlinear cavity which has a uniform gap $\Delta_{\rm S}$ up to a desired photon number $N$ and a smaller gap $\Delta_{\rm S}^{\prime}$ at higher energies, as sketched in Fig.~\ref{fig:Fock}(a). Suppose we couple this cavity to our free-fermion lead with $\smash{\hat{S}_1^-}$ replaced by the photon annihilation operator, and set $\mu_{\rm L}$ such that $\Delta_{\rm S}^{\prime} < \mu_{\rm L} + 2 J_{\rm L} < \Delta_{\rm S}$. Then the cavity will resonantly cool above this level and undergo both heating and cooling below this level, which can stabilize a pronounced peak at $n = N$ if the heating dominates in the latter, as shown in Fig.~\ref{fig:Fock}(b). The sharpness of the peak depends on the heating-to-cooling ratio, which can be found analytically using the transition rates and increases with the nonlinearity $1 - \Delta_{\rm S}^{\prime} / \Delta_{\rm S}$, as expected. For a given nonlinearity, one can amplify heating by modifying the band structure of the lead as in Fig.~\ref{fig:freefermion}(b) or by decreasing $\Delta_{\rm S} / J_{\rm L}$. We quantify these fidelities and the associated preparation times in Appendix \ref{app:Fock_preparation}.

\begin{figure}[tbp]
    \centering
    \includegraphics[width=1\columnwidth]{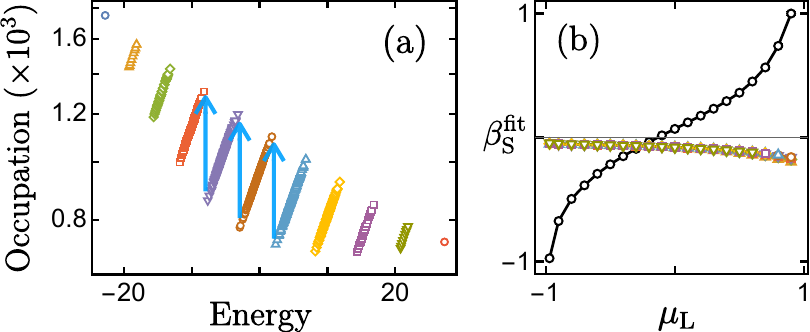}
    \caption{\label{fig:subsector}(a) Steady-state distribution of a $10$-qubit XXZ chain 
    with $J_z = 1$, $J_{\perp} = 0.5$, $B_z = -5$, coupled to the free-fermion lead with $\mu_{\rm L} = -0.1$. The non-overlapping bands represent increasing $S^z$ sectors. Blue arrows show resonant cooling from the bottom of one band to the top of the lower band. (b) Fitted temperature of each band (colored dots) and from the average occupations of all bands (black dots).
    }
\end{figure}

\emph{Global cooling with subsector heating.}|In extended systems one can create athermal states where subspaces heat up internally even though there is an overall cooling. For instance, consider an XXZ chain in a magnetic field such that the energies of the different $S^z$ sectors do not overlap and increase with $S^z$. Then, as Fig.~\ref{fig:subsector} shows, each sector can reach a negative temperature while the overall energy distribution exhibits a crossover from heating to cooling as $\mu_{\rm L}$ is swept across the band. To understand this, note that $\hat{H}_{\rm SL}$ takes the system between successive $S^z$ sectors. The increased population at the top of each sector stems from resonant cooling from the bottom of the next sector [blue arrows in Fig.~\ref{fig:subsector}(a)], as this energy loss is small enough to be compensated by exciting a hole in the lead, $E_{\rm min}(S^z+1) - E_{\rm max}(S^z) < \mu_{\rm L} + 2J_{\rm L}$. On the other hand, the global crossover occurs as the fraction of the occupied lead eigenmodes grows with $\mu_{\rm L}$. For the chosen parameters, global sympathetic cooling does not occur since the gaps between the dominant states are too big, $E_{\rm max}(S^z+1) - E_{\rm max}(S^z) > 4J_{\rm L}$. Note that one can also reverse the scenario (i.e., subsector cooling with overall heating) by heating the lead instead.

\begin{figure}[b]
    \centering
    \includegraphics[width=1\columnwidth]{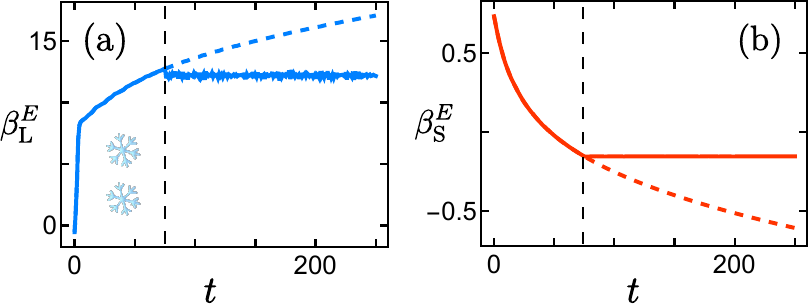}
    \caption{\label{fig:temp_gradient}Evolution of an $8$-site XXZ chain with $J_z = J_{\perp} = 1$ and $B_z = 2$ for the first four ``lead'' sites (a) and $B_z = -2$ for the last four ``system'' sites (b). The lead is cooled with on-site Lindblad operators $\smash{\hat{L}_i = \sqrt{2} \hat{S}_i^+}$. Temperatures obtained from their energies do not change once the cooling is turned off. Dashed curves show the evolution with continuous cooling.
    }
\end{figure}

\emph{Stable temperature gradient.}|For $\mu_{\rm L} < -2 J_{\rm L}$, the coupling $\smash{\hat{c}_1 \hat{S}_1^- + \hat{c}_1^{\dagger} \hat{S}_1^+}$ cannot exchange energy between the lead and the system, and thus cannot eliminate any temperature difference between the two, even if the cooling is turned off. Figure~\ref{fig:temp_gradient} shows that this holds for a more general setting of a uniform Heisenberg chain whose left and right halves are subjected to opposite magnetic fields and the left half is cooled to its ground state $|\!\uparrow \uparrow \uparrow \uparrow \rangle$ using local spin-raising Lindblad operators. The temperatures of the two halves, extracted from their average energies, show that the right half heats up as the left half is cooled and, after the cooling is stopped, the temperature gradient remains. This example demonstrates that the anti-thermalizing dynamics does not rely on the presence of a free-fermion lead nor small $\gamma$ and $\epsilon$, but depends instead on the nature of the coupling.

\begin{figure}[tbp]
    \centering
    \includegraphics[width=1\columnwidth]{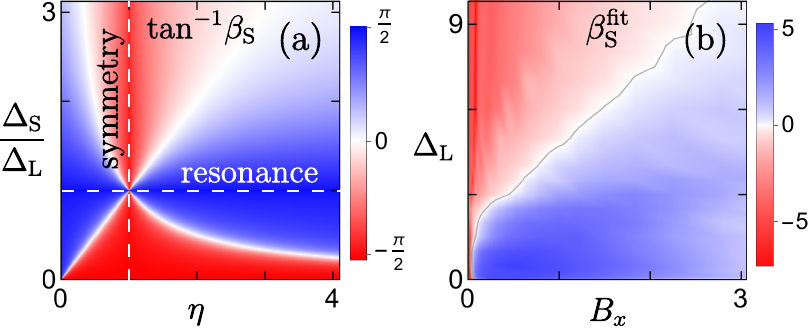}
    \caption{\label{fig:symmetry_breaking}(a) Steady-state temperature for $\smash{\hat{H}_{\rm S}} = \Delta_{\rm S} \smash{\hat{S}^z}$ coupled to the single-qubit lead with $\smash{\hat{H}_{\rm SL}} = (1+\eta) \hat{\sigma}^+ \smash{\hat{S}^-} + (1-\eta) \hat{\sigma}^+ \smash{\hat{S}^+} + \text{h.c.}$, showing maximal heating for $U(1)$ symmetry ($\eta=1$) and maximal cooling on resonance ($\Delta_{\rm S} = \Delta_{\rm L}$). (b) Fitted steady-state temperature of a $15$-qubit XXZ chain with $J_z = 1$, $J_{\perp} = 0.5$, $B_z = -1$, showing a crossover from heating to cooling in the presence of a transverse magnetic field $B_x$.
    }
\end{figure}

\emph{Effect of symmetry breaking.}|A crucial ingredient in the examples we have discussed is a $U(1)$ symmetry (e.g., conservation of total $S^z$) to suppress resonant processes. Let us discuss two ways of breaking this symmetry for a single-qubit lead that is being cooled. First, one can break the symmetry in the coupling to a spin-$S$ system as  $\hat{H}_{\rm SL} = \hat{\sigma}^+ \hat{S}^- + \nu \hat{\sigma}^+ \hat{S}^+ + \text{h.c.}$ The rate equations for this setup gives a thermal steady state for the system with $\beta_{\rm S} = (2 / \Delta_S) \log |\nu (\Delta_{\rm L} + \Delta_{\rm S})/(\Delta_{\rm L} - \Delta_{\rm S})|$ (Appendix \ref{app:detailed_balance}), which is positive for $\nu > |\Delta_{\rm L} - \Delta_{\rm S}| / (\Delta_{\rm L} + \Delta_{\rm S})$. Thus, heating by cooling fades gradually with $\nu$ except when $\Delta_{\rm L} = \Delta_{\rm S}$. The resulting ``phase diagram'' in Fig.~\ref{fig:symmetry_breaking}(a) shows a rich interplay between symmetry and resonance. Second, we can break the symmetry in $\hat{H}_{\rm S}$ by adding a transverse magnetic field $B_x$ to an XXZ chain. As shown in Fig.~\ref{fig:symmetry_breaking}(b), the chain is cooled at large $B_x$, but the onset of cooling grows with $\Delta_{\rm L}$. Both examples show that the anti-thermal behavior survives up to finite symmetry-breaking perturbations, at least for a finite lead.

\emph{Discussion.}|
Our predictions can be tested in existing quantum simulators where one can control on-site potentials and coupling between qubits. For example, tunable spin-exchange interactions arise naturally in superconducting circuits \cite{Blais2021} and Rydberg arrays \cite{Browaeys2020, Morgado2021, Scholl2022}, and have also been realized with cold atoms \cite{Duan2003, Jepsen2020} and trapped ions \cite{Kotibhaskar2024, Monroe2021}. Local dissipation can be realized using lossy resonators \cite{Ma2019} or by measurement and conditional feedback \cite{Wiseman1994}, as discussed below Eq.~\eqref{eq:energy_levels}. 
There are also realizations of anisotropic XY coupling with broken $U(1)$ symmetry using trapped ions \cite{Kotibhaskar2024} and superconducting qubits \cite{Salath2015}, and protocols for the same in Rydberg arrays \cite{Nishad2023}. Note that our basic mechanism can be tested with only two qubits.

One of our results is that one can selectively stabilize mid-spectrum states by choosing a suitably nonuniform level spacing. It would be useful to expand on this in future work and explore whether one can prepare more exotic many-body states of matter or use the local nature of the coupling to selectively populate localized modes in a nonergodic system \cite{DAlessio2016, Abanin2019, Schulz2019}. It is also of fundamental interest to establish what conditions for the anti-thermal dynamics are essential. For instance, we have modeled the dissipation as a nonunitary process that alters the total energy of the system and the lead. Does one recover the same dynamics where the dissipation originates from an external reservoir and the whole setup evolves unitarily? It is also valuable to explore whether the phenomenon of heating by cooling can manifest in a broader class of systems beyond $U(1)$ symmetry.

\begin{acknowledgments}
\emph{Acknowledgments.}|We thank Abhishek Dhar for helpful discussions. SD acknowledges extended research visits to MPIPKS, Dresden, where part of this work was developed.  MH acknowledges financial support from the Deutsche Forschungsgemeinschaft under grant SFB 1143 (project-id 247310070). PM acknowledges funding from the CNRS. JU acknowledges support from the DST Inspire fellowship, Govt.~of India.
\end{acknowledgments}

\appendix

\section{\label{app:detailed_balance}Steady states with detailed balance}

When a system has a finite number of energy levels and the transitions in the Pauli rate equations are limited to neighboring levels, it follows that the steady state has detailed balance. To see this, consider general rate equations of the form
\begin{equation}
    \frac{{\rm d}p_i}{{\rm d}t} = 
    p_{i+1} \Gamma_{i+1 \to i} 
    + p_{i-1} \Gamma_{i-1 \to i} 
    - p_i (\Gamma_{i \to i+1} + \Gamma_{i \to i-1}) \;,
\end{equation}
where $p_i$ is the population of the $i$-th level and $\Gamma$'s are the transition rates. In the steady state,
\begin{equation}
    p_{i+1} = [p_i (\Gamma_{i \to i+1} + \Gamma_{i \to i-1}) - p_{i-1} \Gamma_{i-1 \to i}]/\Gamma_{i+1 \to i} \;.
    \label{eq:rate_eqn_det_bal}
\end{equation}
For the lowest two levels ($i=1,2$), one finds (using $p_0=\Gamma_{1 \to 0} = 0$), $p_2 \Gamma_{2 \to 1} = p_1 \Gamma_{1 \to 2}$. However, from Eq.~\eqref{eq:rate_eqn_det_bal}, $p_{i-1} \Gamma_{i-1 \to i} = p_i \Gamma_{i \to i-1}$ implies $p_{i} \Gamma_{i \to i+1} = p_{i+1} \Gamma_{i+1 \to i}$. Hence, we have pairwise equilibrium. Below we work out three examples where such a situation holds.

\subsubsection{\label{app:finite_temp}1. Single-site, finite-temperature lead}

First, consider the system $\hat{H}_{\rm S} = \Delta_{\rm S} \hat{S}^z$ coupled to a lead qubit $\hat{H}_{\rm L} = -\Delta_{\rm L} \hat{\sigma}^z$ with $\hat{H}_{\rm SL} = \hat{\sigma}^+ \hat{S}^- + \hat{\sigma}^- \hat{S}^+$. The energy levels of the system are given by $E_m = \Delta_{\rm S} m$, where $m \in \{-S, -S+1, \dots, S\}$. The lead is kept at an inverse temperature $\beta_{\rm L}$ using two Lindblad operators $\hat{L}_{\pm} = \sqrt{\gamma_{\pm}} \hat{\sigma}^{\mp}$ with $\gamma_+ / \gamma_- = e^{-\beta_{\rm L} \Delta_{\rm L}}$. Then the leading-order transition rates from Eq.~\eqref{eq:single-qubit_rates} generalize to
\begin{align}
    \nonumber & \Gamma_{i \to j} = 
    \sum_{\sigma = \uparrow, \downarrow} P_{\sigma}
    \sum_{k = \pm} 
    | {}_{\epsilon} \langle \sigma, E_j | \hat{L}_k | \sigma, E_i \rangle_{\epsilon} |^2 
    \\
    &\!\! = P_{\uparrow} \!\left( \gamma_- |c^{\uparrow, \downarrow}_{i,j} |^2 + \gamma_+ |c^{\uparrow, \downarrow}_{j,i} |^2 \right) 
    + P_{\downarrow} \!\left( \gamma_- |c^{\downarrow, \uparrow}_{j,i} |^2 + \gamma_+ |c^{\downarrow, \uparrow}_{i,j} |^2 \right) \!,
    \label{eq:rate_finite_temp}
\end{align}
where $P_{\sigma}$ are the thermal occupations of the lead and $c^{\sigma, \bar{\sigma}}_{i,j}$ (with $\bar{\uparrow} \coloneqq \downarrow$, $\bar{\downarrow} \coloneqq \uparrow$) are expansion coefficients for the perturbed eigenstates, $|\sigma, E_i \rangle_{\epsilon} = \sum_j c^{\sigma, \sigma}_{i,j} |\sigma, E_j\rangle + c^{\sigma, \bar{\sigma}}_{i,j} |\bar{\sigma}, E_j\rangle$. From first-order perturbation theory,
\begin{align}
    \nonumber c^{\uparrow, \downarrow}_{i,j} &= 
    \frac{\langle \downarrow, E_j | \epsilon \hat{H}_{\rm SL} | \!\uparrow, E_i \rangle}{E_i - E_j - \Delta_{\rm L}} = \epsilon \delta_{j, i+1} \frac{[S(S\!+\!1) - i(i\!+\!1)]^{\frac{1}{2}}}{-\Delta_{\rm S} - \Delta_{\rm L}} ,
    \\
    c^{\downarrow, \uparrow}_{i,j} &= 
    \frac{\langle \uparrow, E_j | \epsilon \hat{H}_{\rm SL} | \!\downarrow, E_i \rangle}{E_i - E_j + \Delta_{\rm L}} = \epsilon \delta_{i, j+1} \frac{[S(S\!+\!1) - j(j\!+\!1)]^{\frac{1}{2}}}{\Delta_{\rm S} + \Delta_{\rm L}} .
    \label{eq:cij}
\end{align}
Using these in Eq.~\eqref{eq:rate_finite_temp}, one finds that the transitions connect only neighboring energy levels of the system, with $\Gamma_{m \to m+1} / \Gamma_{m+1 \to m} = e^{\beta_{\rm L} \Delta_{\rm L}}$. From detailed balance, it follows that $p_{m+1} / p_{m} = e^{\beta_{\rm L} \Delta_{\rm L}}$, which is a thermal state with inverse temperature $\beta_{\rm S} = -(\Delta_{\rm L} / \Delta_{\rm S}) \beta_{\rm L}$.

\subsubsection{\label{app:symmetry_breaking}2. Single-site lead with broken $U(1)$ symmetry}

Next, consider the same setup but where the coupling breaks $U(1)$ symmetry, $\hat{H}_{\rm SL} = \hat{\sigma}^+ \hat{S}^- + \nu \hat{\sigma}^+ \hat{S}^+ + \text{h.c.}$, and the lead is cooled with the Lindblad operator $\hat{L} = \sqrt{\gamma} \hat{\sigma}^+$. From Eqs.~\eqref{eq:rate_finite_temp} and \eqref{eq:cij}, $\Gamma_{i \to j} = \gamma \smash{|c^{\uparrow, \downarrow}_{i,j}|^2}$, with
\begin{equation}
    \frac{c^{\uparrow, \downarrow}_{i,j}}{\epsilon} = 
    \left(\! \frac{\nu\delta_{i,j+1}}{\Delta_{\rm S} - \Delta_{\rm L}} - \frac{\delta_{j,i+1}}{\Delta_{\rm S} + \Delta_{\rm L}} \right) 
    [S(S\!+\!1) - i_< (i_<\!+\!1)]^{\frac{1}{2}} ,
\end{equation}
where $i_<$ is the smaller of $i$ and $j$. Thus, again the transitions occur between neighboring levels, yielding a steady state with $p_m/p_{m+1} = [\nu (\Delta_{\rm S} + \Delta_{\rm L})/(\Delta_{\rm S} - \Delta_{\rm L})]^2$, which corresponds to $\beta_{\rm S} = (2 / \Delta_S) \log |\nu (\Delta_{\rm L} + \Delta_{\rm S})/(\Delta_{\rm L} - \Delta_{\rm S})|$.

\subsubsection{\label{app:free_fermion}3. Free-fermion lead}

Finally, consider a free-fermion lead with dispersion $\varepsilon_{\rm L}(k)$ and eigenmodes $\hat{\chi}_k = [2/(l\!+\!1)]^{\frac{1}{2}}\sum_i \alpha_{i}(k) \hat{c}_i$, where $l$ is the number of sites and $\hat{c}_i$ annihilates a fermion at site $i$. Site $1$ of the lead is coupled to the system as $\hat{H}_{\rm SL} = \hat{c}_1 \hat{S}^- + \hat{c}_1^{\dagger} \hat{S}^+$ and the lead is being cooled to its ground state $|g\rangle$ where all modes below the chemical potential $\mu_{\rm L}$ are occupied and the rest are empty. The cooling is achieved with Lindblad operators $\hat{L}_k = \sqrt{\gamma} \smash{\hat{\chi}_k^{\dagger}}$ for $\varepsilon_{\rm L}(k) < \mu_{\rm L}$ and $\smash{\hat{L}_k} = \sqrt{\gamma} \hat{\chi}_k$ for $\varepsilon_{\rm L}(k) > \mu_{\rm L}$. To find the transition rates in the system for $\epsilon \ll \gamma$, it suffices to consider only single-particle excitations of the lead, $|k\rangle \coloneqq \gamma^{-1/2}\smash{\hat{L}_k^{\dagger}} |g\rangle$, in the perturbed energy eigenstates
\begin{equation}
    |g, E_i \rangle_{\epsilon} = \sum\nolimits_j c^{g,g}_{i,j} |g, E_j \rangle + \sum\nolimits_{k,j} c^{g,k}_{i,j} |k, E_j \rangle + \cdots ,
\end{equation}
where, to $O(\epsilon)$, $c^{g,g}_{i,j} = \delta_{i,j}$ and 
\begin{equation}
    c^{g,k}_{i,j} = \frac{\langle k, E_j | \epsilon \hat{H}_{\rm SL} | g, E_i \rangle}{E_i - E_j - |\varepsilon_{\rm L}(k) - \mu_{\rm L}|} \;.
    \label{eq:cgkij}
\end{equation}
Using $\hat{c}_i = \sum_k \alpha_i^*(k) \hat{\chi}_k$ yields
\begin{equation}
    \langle k, E_j | \hat{H}_{\rm SL} | g, E_i \rangle = 
    \begin{cases}
        \alpha_1^*(k) \langle E_j | \hat{S}^- | E_i \rangle & \text{if } \varepsilon_{\rm L}(k) < \mu_{\rm L} \\
        \alpha_1(k) \langle E_j | \hat{S}^+ | E_i \rangle & \text{if } \varepsilon_{\rm L}(k) > \mu_{\rm L}
    \end{cases}
    \;.
    \label{eq:coupling_matrix_element}
\end{equation}
As before, the leading-order transition rates are given by 
\begin{equation}
    \Gamma_{i \to j} 
    = \sum\nolimits_k | {}_{\epsilon} \langle g, E_j | \hat{L}_k | g, E_i \rangle_{\epsilon} |^2 
    = \gamma \sum\nolimits_k |c^{g,k}_{i,j}|^2 .
\end{equation}
Using Eqs.~\eqref{eq:cgkij} and \eqref{eq:coupling_matrix_element} one finds, for an infinite lead,
\begin{equation}
    \frac{\Gamma_{i \rightarrow j}}{\gamma \epsilon^2} = \frac{2}{\pi}
    \sum_{p=\pm} \int_{p(\varepsilon_{\rm L}(k) - \mu_{\rm L}) > 0} \!
    \frac{{\rm d}k \; |\alpha_1(k)|^2 \; |\langle E_j | \hat{S}^p | E_i \rangle|^2}{(E_i - E_j - |\varepsilon_{\rm L}(k) - \mu_{\rm L}|)^2} .
\end{equation}
For a ``tight-binding'' lead with nearest-neighbor hopping $J_{\rm L}$, $\varepsilon_{\rm L}(k) = -2 J_{\rm L} \cos k$ and $\alpha_1(k) = \sin k$. Then the rates can be found analytically as
\begin{equation}
    \tilde{\Gamma}_{i \to j} = 
    \sum_{p=\pm} |\langle E_j | \hat{S}^p | E_i \rangle|^2 \; I \!\left(\! \frac{E_i - E_j + p \mu_{\rm L}}{2 J_{\rm L}}, \cos^{-1} \frac{p\mu_{\rm L}}{2 J_{\rm L}} \!\right) \!,
    \label{eq:tight_binding_rates}
\end{equation}
where $\mu_{\rm L} \in [-2 J_{\rm L}, 2 J_{\rm L}]$, $\tilde{\Gamma}_{i \to j} \coloneqq (2\pi/\gamma) (J_{\rm L} / \epsilon)^2 \Gamma_{i \to j}$ and 
\begin{align}
    \nonumber I(x,\theta) \coloneqq &\;
    \int_0^{\theta} \! \frac{{\rm d}k \;\sin^2 k}{(x - \cos k)^2} \\
    =&\;  \frac{\sin\theta}{\cos\theta - x} -\theta
    + \frac{2x}{\sqrt{x^2-1}} \tan^{-1} \frac{(1+x) \tan \frac{\theta}{2}}{\sqrt{x^2-1}} \;.
    \label{eq:tight_binding_integral}
\end{align}
For $\hat{H}_{\rm S} = \Delta_{\rm S} \hat{S}^z$, Eq.~\eqref{eq:tight_binding_rates} gives transitions only between adjacent levels with $\Gamma_{m \to m+1} / \Gamma_{m+1 \to m} = I_+ / I_-$, where $I_{\pm} \coloneqq I \big(\!\pm\!(\mu_{\rm L} - \Delta_{\rm S})/(2 J_{\rm L}), \; \cos^{-1}(\pm \mu_{\rm L}/(2 J_{\rm L})) \big)$. This leads to a thermal state with $\beta_{\rm S} = \log(I_- / I_+)/\Delta_{\rm S}$, which is plotted in Fig.~\ref{fig:freefermion}(c) for $J_{\rm L} = 0.5$ and $\Delta_{\rm S} = 1.5$. For $\mu_{\rm L} + 2J_{\rm L} > \Delta_{\rm S}$, $I_-$ diverges due to resonance, which can be regularized by including higher-order terms in $\gamma$ \cite{Reichental2018}.

\section{\label{app:third_quantization}Exact results using third quantization}

\begin{figure}[tbp]
    \centering
    \includegraphics[width=1\columnwidth]{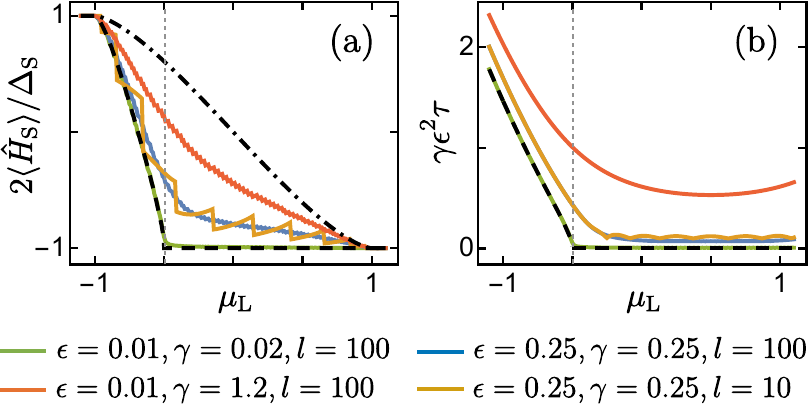}
    \caption{\label{fig:third_quantization}(a) Steady-state energy and (b) relaxation time for a single-qubit system with gap $\Delta_{\rm S} = 0.5$ coupled to the free-fermion lead with $J_{\rm L} = 0.5$ for various coupling strengths $\epsilon$, cooling rates $\gamma$, and lead sizes $l$, from third quantization. Dashed curves are from the rate equations [Eq.~\eqref{eq:tight_binding_rates}] and the dot-dashed curve in (a) is the quantum Zeno limit [Eq.~\eqref{eq:quantum_zeno}].
    }
\end{figure}

\begin{figure}[tbp]
    \centering
    \includegraphics[width=1\columnwidth]{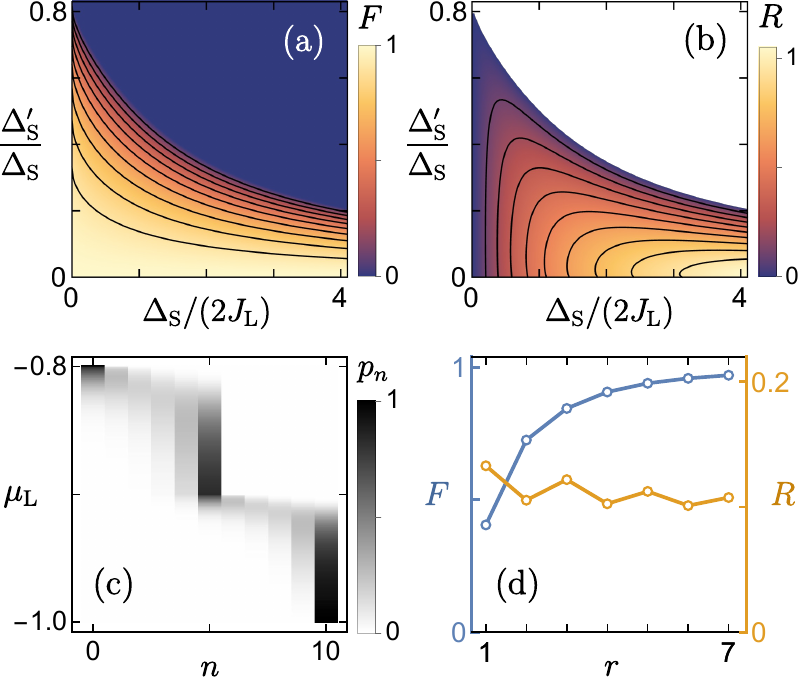}
    \caption{\label{fig:Fock_details}(a) Fidelity and (b) relaxation rate $R = \pi\Delta_{\rm S}^2 / (2 \gamma \epsilon^2 \tau)$ for the preparation of a Fock state in a nonlinear cavity with gaps $\Delta_{\rm S}$ and $\Delta_{\rm S}^{\prime}$, as in Fig.~\ref{fig:Fock}(a), coupled to the free-fermion lead with nearest-neighbor hops. (c) Steady-state distribution for a lead with longer-range hopping ($r=3$), as sketched in Fig.~\ref{fig:freefermion}(b), with $J_{\rm L} = 0.5$, $\Delta_{\rm S} = 0.2$, $\Delta_{\rm S}^{\prime} = 0.1$. (d) Fidelity and relaxation rate vs hopping range for $J_{\rm L} = \Delta_{\rm S} = 2 \Delta_{\rm S}^{\prime}$.
    }
\end{figure}

When the system and the lead are both free fermionic, one can solve the dynamics exactly by writing the Liouvillian as a quadratic form in fermionic superoperators, called third quantization \cite{Prosen2008}. This reduces the complexity from exponential to linear in the number of sites, allowing exact simulations of large setups. Here we simulate a single-qubit system, $\hat{H}_{\rm S} = \Delta_{\rm S} \hat{S}^z$, coupled to the free-fermion lead. As shown in Fig.~\ref{fig:third_quantization}(a), the system is maximally heated for $\mu_{\rm L} < -2J_{\rm L}$ and maximally cooled for $\mu_{\rm L} > 2J_{\rm L}$ for all coupling strengths $\epsilon$ and cooling rates $\gamma$. For small $\epsilon$ and $\gamma$, we recover the predictions of the rate equations in the last section (dashed curve), which produces a sharp threshold at the onset of sympathetic cooling, $\mu_{\rm L} + 2 J_{\rm L} = \Delta_{\rm S}$. Away from this limit, the resonance is broadened and one obtains a smooth crossover. For a finite lead, the modes cross below $\mu_{\rm L}$ one at a time, producing a sawtooth-like variation. For $\gamma \gg J_{\rm L}, \Delta_{\rm S}$, the threshold is washed out and the crossover is uniform across the band (dot-dashed curve). In this quantum Zeno limit, one can trace out the lead following Ref.~\cite{Popkov2018} and obtain an effective dynamics of the system with two Lindblad operators $\hat{L}_{\pm} = \sqrt{\Gamma_{\pm}} \hat{S}^{\pm}$, where
\begin{equation}
    \Gamma_{\pm} = \frac{4\epsilon^2}{\pi\gamma} 
    \left[ \cos^{-1} \!\big(\!\pm \tilde{\mu}_{\rm L} \big) \mp \tilde{\mu}_{\rm L} \sqrt{1 - \tilde{\mu}_{\rm L}^2} \right] ,
    \label{eq:quantum_zeno}
\end{equation}
$\tilde{\mu} \coloneqq \mu_{\rm L} / (2 J_{\rm L})$. This gives $2 \langle \hat{S}^z \rangle = (\Gamma_+ - \Gamma_-) / (\Gamma_+ + \Gamma_-)$ in the steady state and a relaxation time $\tau = 2/(\Gamma_+ + \Gamma_-) = \gamma/(2\epsilon^2)$, independent of $\mu_{\rm L}$. On the other hand, for small $\gamma$ and $\epsilon$, $\tau \sim J_{\rm L}^2/[\gamma \epsilon^2 (I_+ + I_-)]$ [see Eq.~\eqref{eq:tight_binding_rates}], which depends strongly on $\tilde{\mu}_{\rm L}$ and $\Delta_{\rm S} / J_{\rm L}$. These different scalings are reproduced by exact numerics in their respective limits [see Fig.~\ref{fig:third_quantization}(b)]. In both cases, $\tau$ does not vary significantly with the lead size.

\section{\label{app:Fock_preparation}Fidelity and timescales for preparation of a mid-spectrum (Fock) state}

In Fig.~\ref{fig:Fock} we showed that the free-fermion lead can be used to produce a mid-spectrum state, $n=N$, in a non-linear cavity which has a larger gap $\Delta_{\rm S}$ for $n < N$ and a smaller gap $\Delta_{\rm S}^{\prime}$ for $n > N$. The fidelity $F$ is highest for $\mu_{\rm L} = -2J_{\rm L} + \Delta_{\rm S}^{\prime}$, when the higher levels are resonantly cooled and the heating-to-cooling ratio for $n < N$ is maximized. Using detailed balance gives $p_n / p_{n+1} = I_- / I_+$, where $I_{\pm}(\Delta_{\rm S}^{\prime} / \Delta_{\rm S}, \Delta_{\rm S} / J_{\rm L})$ are defined below Eq.~\eqref{eq:tight_binding_integral}. Thus, for $I_- < I_+$ and $N \gg 1$, $F \coloneqq p_N = 1 - I_- / I_+$. Figure~\ref{fig:Fock_details}(a) shows that $F$ increases as either $\Delta_{\rm S}^{\prime} / \Delta_{\rm S}$ or $\Delta_{\rm S} / J_{\rm L}$ is reduced. The preparation time $\tau$ can be found from the gap of the rate matrix, which for $N \gg 1$ yields $\tau = \pi\Delta_{\rm S}^2 / (2 \gamma \epsilon^2 R)$ with $R \coloneqq (I_+ - I_-) \Delta_{\rm S}^2 / (4 J_{\rm L}^2)$. For a given nonlinearity $\Delta_{\rm S}^{\prime} / \Delta_{\rm S}$, $\tau$ is minimized for an optimal choice of $\Delta_{\rm S} / J_{\rm L}$, as shown in Fig.~\ref{fig:Fock_details}(b). Figures~\ref{fig:Fock_details}(c) and ~\ref{fig:Fock_details}(d) show that one can also enhance the fidelity without compromising speed by engineering the lead dispersion $\varepsilon_{\rm L}(k) = A_r - B_r \sum_{m=1}^r \cos (m k) / m$, where $A_r$ and $B_r$ are chosen to satisfy $\varepsilon_{\rm L}(\pi) = -\varepsilon_{\rm L}(0) = 2 J_{\rm L}$. The higher fidelity stems from more heating due to fewer occupied modes at a given $\mu_{\rm L}$, as sketched in Fig.~\ref{fig:freefermion}(b).

\end{document}